\documentclass[english,aps,manuscript,preprint,showpacs]{revtex4}
\usepackage[T1]{fontenc}
\usepackage[latin9]{inputenc}
\usepackage{textcomp}
\usepackage{relsize}
\usepackage{amsmath}
\usepackage{graphicx}
\usepackage{amssymb}

\makeatletter

\DeclareRobustCommand{\lyxmathsym}[1]{\ifmmode\begingroup\def\b@ld{bold}
  \def\rmorbf##1{\ifx\math@version\b@ld\textbf{##1}\else\textrm{##1}\fi}
  \mathchoice{\hbox{\rmorbf{#1}}}{\hbox{\rmorbf{#1}}}
  {\hbox{\smaller[2]\rmorbf{#1}}}{\hbox{\smaller[3]\rmorbf{#1}}}
  \endgroup\else#1\fi}

\makeatother

\makeatother

\usepackage{babel}

\begin{document}

\title{Experimental violation of a Bell's inequality in time with weak measurement}

\author{Agustin~Palacios-Laloy$^{1}$}

\author{François~Mallet$^{1}$}

\author{François~Nguyen$^{1}$}

\author{Patrice~Bertet$^{1}$$^{\text{*}}$}

\author{Denis~Vion$^{1}$}

\author{Daniel~Esteve$^{1}$}

\author{Alexander~Korotkov$^{2}$}

\affiliation{$^{1}$Quantronics Group, Service de Physique de l'État Condensé
(CNRS URA 2464), DSM/IRAMIS/SPEC, CEA-Saclay, 91191 Gif-sur-Yvette
Cedex, France}

\affiliation{$^{2}$Deparment of Electrical Engineering, University of California,
Riverside, CA 92521-0204, USA}

\pacs{74.50,03.65,82.25}
\begin{abstract}
The violation of J. Bell's inequality with two entangled and spatially
separated quantum two-level systems (TLS) is often considered as the
most prominent demonstration that nature does not obey {}``local
realism''. Under different but related assumptions of {}``macrorealism'',
plausible for macroscopic systems, Leggett and Garg derived a similar
inequality for a \emph{single} degree of freedom undergoing coherent
oscillations and being measured at successive times. Such a {}``Bell's
inequality in time'', which should be violated by a quantum TLS,
is tested here. In this work, the TLS is a superconducting quantum
circuit whose Rabi oscillations are continuously driven while it is
continuously and weakly measured. The time correlations present at
the detector output agree with quantum-mechanical predictions and
violate the inequality by 5 standard deviations. 
\end{abstract}
\maketitle

\subsection*{Introduction}

The violation of J. Bell's inequality \cite{Bell,CHSH} is the most
prominent example of a situation where the predictions of quantum
mechanics are incompatible with a large class of classical theories.
In the early 1980s, Aspect and coworkers \cite{Aspect} brought an
experimental proof of this violation using pairs of spatially-separated
polarization-entangled photons. By demonstrating an excess of correlations
between the polarizations measured on the two photons of a pair, they
ruled out descriptions of nature satisfying the very general conditions
known as local realism. This striking finding also contributed to
transform the so-called quantum weirdness into a useful resource for
information processing. Shortly after, quantum cryptography protocols
and quantum algorithms exploiting entanglement were indeed proposed
\cite{Chuang}. Following a reasoning similar to that of Bell, Leggett
and Garg derived in 1985 an inequality that can be seen as a {}``Bell's
inequality in time'', which applies to any \textit{single} macroscopic
system measured at successive times \cite{GargLeggett} and fullfiling
the assumptions of macrorealism: (A1) the system is always in one
of its macroscopically distinguishable states, and (A2) this state
can be measured in a non invasive way, i.e. without perturbing the
subsequent dynamics of the system. Quantum mechanics however contradicts
both assumptions, which can lead to an excess of correlations between
subsequent measurements and to a violation of this inequality. Ruskov
and coworkers \cite{Ruskov} then adapted the inequality to the situation
where a two-level systems (TLS) is continuously and weakly monitored
during its coherent oscillations. Using such a weak monitoring, we
report here an experimental test of a Bell's inequality in time (see
also the recent works \cite{Goggin,Xu}), yielding results in excellent
agreement with simple quantum-mechanical predictions and in contradiction
with a large class of macrorealistic models.

\subsection*{Bell's inequalities in space and in time}

\begin{figure}
\includegraphics[width=8.8cm]{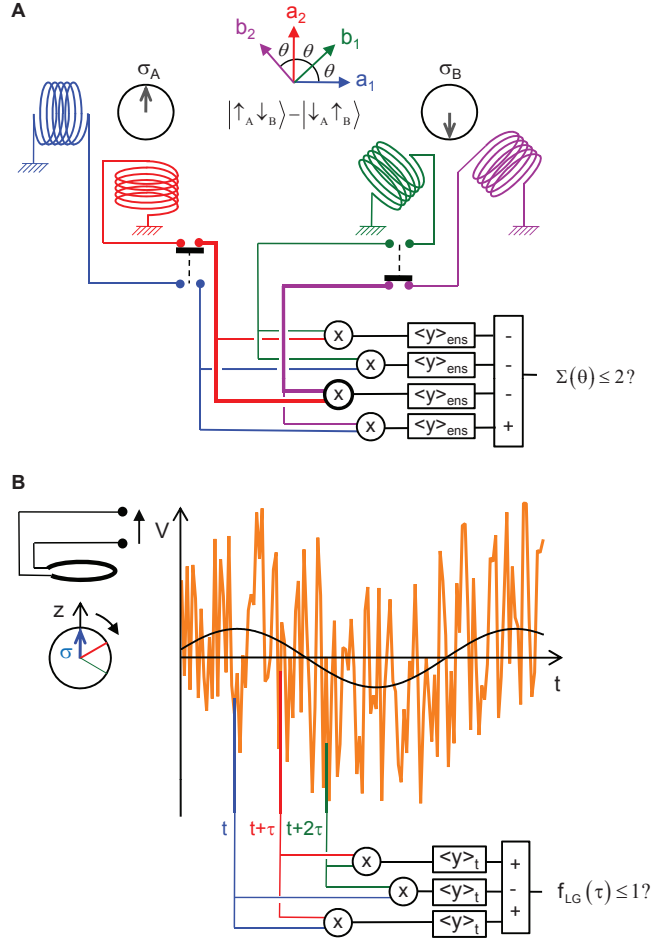}

\caption{Comparison between two thought experiments which test the usual CHSH
Bell's inequality and the Bell's inequality in time. (A) CHSH inequality:
two maximally entangled spins $\sigma_{A}$ and $\sigma_{B}$ are
sent to two spatially separated observers A and B. Each of the observers
measures with pick-up coils his spin along one of two possible directions
($a_{1}$ and $a_{2}$ for A, and $b_{1}$ and $b_{2}$ for B ); the
four directions make angles $\theta$ as depicted. By repeating this
experiment on a statistical ensemble, a linear combination $\Sigma$
of the four possible correlators between measurements on a spin pair
is computed. Local realism requires $-2\leq\Sigma\leq2$, while quantum
mechanics predicts $\Sigma=2\,\sqrt{2}$ for $\theta=45\lyxmathsym{\textdegree}$.
(B) Bell's inequality in time with weak measurement: a single spin
$\sigma$ undergoing coherent oscillations at frequency $\omega_{R}$
is continuously measured with a pick-up coil coupled to it so weakly
that the time for a complete projective measurement would be much
longer than the period of oscillations $T_{R}=2\pi/\omega_{R}$. From
the noisy time trace recorded in the steady state, one computes a
linear combination $f_{LG}$ of the three time-averaged-correlators
between the readout outcomes at three times separated by $\tau$.
Macrorealism requires $f_{LG}\leq1$ for any $\tau$, while quantum
mechanics predicts $f_{LG}=1.5$ at $\tau=T_{R}/6$.}

\label{fig1} 
\end{figure}

We start by briefly recalling the experimental protocol of the usual
CHSH test \cite{CHSH} of Bell's inequalities (see Fig.~\ref{fig1}a).
It consists in identically preparing many times a pair of quantum
TLS in a maximally entangled state such as $\left|\psi^{-}\right\rangle =(\left|\uparrow\downarrow\right\rangle -\left|\downarrow\uparrow\right\rangle )/\sqrt{2}$.
Each member of the pair is then distributed to two observers A and
B, who perform projective measurements of the TLS spin $\sigma_{i}^{A,B}=\pm1$
along one of two directions $a_{i}$ ($i=1,2$) for A and $b_{i}$
for B, with these directions forming angles $(a_{1},b_{1})=(b_{1},a_{2})=(a_{2},b_{2})\equiv\theta$
as shown in Fig.~\ref{fig1}a. The two observers then combine all
their measurements to compute the Bell sum $\Sigma(\theta)=-K_{11}+K_{12}-K_{22}-K_{21}$
of the correlators $K_{ij}(\theta)=\langle\sigma_{i}^{A}\sigma_{j}^{B}\rangle$.
The Bell's theorem, based on a simple statistical argument, states
that according to all local realistic theories \begin{equation}
-2\leq\Sigma(\theta)\leq2.\label{eq:Bell}\end{equation}
However, standard quantum mechanics predicts that this inequality
is violated, with a maximum violation $\Sigma(\theta=\pi/4)=2\sqrt{2}$.
Many experimental tests, and in particular those performed by A. Aspect
\cite{Aspect} have verified this violation \cite{Martinis,Rob Bell}. 

While quantum entanglement between two spatially separated TLS is
at the heart of the previous violation, Leggett and Garg proposed
a similar inequality \cite{GargLeggett} holding for a \emph{single
}degree of freedom $-1\leq z(t)\leq1$ fulfilling the assumptions
of macrorealism (z(t) defined at any time, and measurable with no
perturbation). Using a simple arithmetic argument \textit{à la Bell},
they showed that

\begin{equation}
z(t_{0})z(t_{1})+z(t_{1})z(t_{2})-z(t_{0})z(t_{2})\leq1\label{eq:z(t)}\end{equation}
 for all $\left\{ t_{i}\right\} $. Consequently, an observer measuring
$z$ on many identical systems, either at $t_{0}$ and $t_{1}=t_{0}+\tau$,
or at $t_{0}$ and $t_{2}=t_{0}+2\tau$, or at $t_{1}$ and $t_{2}$
should find ensemble-averaged correlators $K_{ij}(t_{0},\tau)=\langle z(t_{i})z(t_{j})\rangle$
(for $i,j=0,1,2$, with $i<j$) satisfying the Leggett-Garg's inequality:
\begin{equation}
f_{LG}(t_{0},\tau)\equiv K_{01}+K_{12}-K_{02}\leq1.\label{eq:LG-1}\end{equation}
 Quantum mechanics on the other hand predicts that, applied to the
case of a quantum TLS undergoing coherent oscillations at frequency
$\omega_{R}$ , this inequality is violated for well-chosen values
of $\tau$, with maximum violation $f_{LG}(t_{0},\tau=\pi/3\omega_{R})=1.5$
independent of $t_{0}$. Here, the delay $\tau$ between successive
measurements plays the role of the angle $\theta$ between the measurement
directions in the Bell's inequality (\ref{eq:Bell}), justifying the
nickname {}``Bell's inequality in time''. The excess of correlations
predicted by quantum mechanics, compared to the macrorealistic case,
can be interpreted as resulting from the projection of the TLS state
on a $\sigma_{z}$ eigenstate induced by the first measurement.

As shown in \cite{Ruskov}, the very same conclusions also hold if
the TLS undergoing coherent oscillations is continuously and weakly
monitored along $\sigma_{z}$ (see Fig.~1B) instead of being projectively
measured at well-defined times. The detector now delivers an output
signal $V(t)=(\delta V/2)z(t)+\xi(t)$ proportional to \emph{z(t)}
with some additional noise $\xi(t)$. Macrorealism implies that the
dynamics of the system at time $t+\tau$ is fully uncorrelated with
the detector noise at time $t$ so that $\langle\xi(t)z(t+\tau)\rangle_{t}=0$.
The detector's output correlation function $K(\tau)=\langle V(t)V(t+\tau)\rangle_{t}/(\delta V/2)^{2}$
is then simply equal to $\langle z(t)z(t+\tau)\rangle_{t}$. By averaging
inequality (\ref{eq:z(t)}) over $t_{0}$ in the steady-state, the
Bell's inequality in time (\ref{eq:LG-1}) becomes \begin{equation}
f_{LG}(\tau)\equiv2K(\tau)-K(2\tau)\leq1,\label{eq:LG}\end{equation}
 and should be violated by a quantum TLS in the very same way as discussed
above. Here the violation is however not due to a strong projection
of the TLS wavefunction induced by measurements at well-defined times
of its evolution, but rather to the continuous partial projection
caused by the measurement during the TLS coherent evolution, which
reinforces correlations between the detector output at successive
times.

\subsection*{Experimental setup}

\begin{figure}[tbh]
\includegraphics[width=8.8cm]{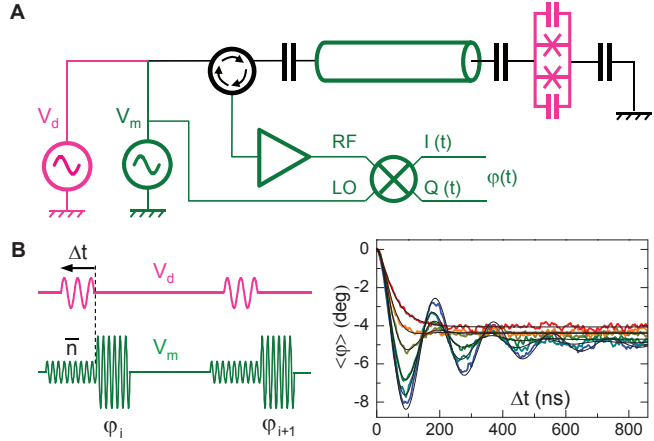}

\caption{(A) Experimental implementation of the thought experiment in Fig.~\ref{fig1}B)
with a quantum electrical circuit. The spin (or TLS) is a Cooper pair
box of the transmon type (magenta) capacitively coupled to a microwave
resonator (cavity sketched as a green coaxial cable). Two microwave
sources $V_{d}$ and $V_{m}$ are used to drive and measure the transmon
at $\omega_{ge}$ and $\omega_{c}$, respectively. The reflected microwave
at $\omega_{c}$ is routed through a circulator to a cryogenic amplifier
followed by an I-Q demodulator. The time dependent phase $\varphi(t)$
measured by the demodulator carries the information about the TLS
state. (B) Measurement-induced dephasing of the TLS as a function
of the measurement strength, i.e. the amplitude of $V_{m}$ or the
mean photon number $\bar{n}$ in the cavity. Measured ensemble-averaged
Rabi oscillations $\langle\varphi(\Delta t)\rangle$ in presence of
$\bar{n}=0$, $1$, $2$, $5$, $10$ and $20$ photons (from blue
to red curve), showing the transition from weak to strong measurement.
Each curve is fitted to a solution of Bloch equations (thin black
lines), in quantitative agreement with expected measurement-induced
dephasing rates (see online supplementary information\ref{sub:Measurement-induced-dephasing}).}

\label{fig2} 
\end{figure}

Our experimental setup (see Fig.~\ref{fig2}A and supplementary information\ref{sub:Sample-fabrication})
for probing inequality (\ref{eq:LG}) closely implements the proposal
discussed above, making use of the possibilities offered by the so-called
circuit quantum electrodynamics (circuit-QED) architecture \cite{Blais,Wallraff}
where a superconducting artificial TLS is coupled to a superconducting
coplanar waveguide resonator. The TLS consists here of the two lowest
energy states $g$ and $e$ of a modified Cooper-pair box of the transmon
type \cite{transmon_th,transmon_exp}. These two states can be regarded
as {}``macroscopically distinguishable'' because the dipole moment
of the $g-e$ transition is of the order of $10^{4}$ atomic units.
On the other hand, the only degree of freedom of this system is the
phase difference between the superconducting order parameters on both
sides of the Josephson junction forming the Cooper-pair box, conjugate
to the number of Cooper pairs passed through the junction; this phase
is a collective variable whose degree of macroscopicity is still under
debate \cite{Macro,Leggett2008}.

The TLS transition frequency is $\omega_{ge}/2\pi=5.304\,$GHz, below
the resonance frequency $\omega_{c}/2\pi=5.796\,$GHz of the resonator
to which it is capacitively coupled for its measurement. Two microwave
sources $V_{d}$ and $V_{m}$ drive and measure the TLS at frequencies
$\omega_{ge}$ and $\omega_{c}$, respectively. In order to continuously
monitor the induced Rabi oscillations up to a few tens of MHz, we
implement a resonator bandwidth of $\kappa/2\pi=30.3\pm0.8\,$MHz
(quality factor $191\pm5$) by designing the appropriate resonator
input capacitance \cite{Wallraff}. With these parameters, the TLS
is sufficiently detuned from the resonator for their interaction to
be well described by the so-called dispersive Hamiltonian $\hat{H}=\hbar\chi\hat{n}\hat{\sigma_{z}}$
\cite{Blais}, with $\hat{n}$ the photon number inside the readout
mode and $\chi$ the dispersive coupling constant. The resonator frequency
is thus shifted by $\pm\chi/2\pi=\pm1.75[-0.11/+0.14]\,$MHz depending
on the TLS state (see supplementary information\ref{sub:Sample-parameters}).
The phase $\varphi$ of a microwave signal at $\omega_{c}$ therefore
acquires a TLS state-dependent shift after being reflected by the
resonator, and provides a non-destructive readout of the TLS as demonstrated
in numerous experiments \cite{Blais,Gambetta,Wallraff}. In our setup,
the reflected signal is routed through a circulator to a cryogenic
amplifier and is then measured by homodyne detection at room temperature,
yielding the two field quadratures $I(t)$ and $Q(t)$. These time
traces provide a continuous measurement of the TLS with a strength
proportional to the signal input power, and thus to the intra-resonator
average photon number $\overline{n}$. In a fully quantum-mechanical
description of the measurement process using the quantum trajectory
formalism \cite{Gambetta}, each quadrature can be written $X(t)=\bar{X}+(\delta X/2)\left\langle \hat{\sigma}_{z}\right\rangle _{c}(t)+\xi_{0}(t)$,
where $\left\langle \hat{\sigma}_{z}\right\rangle _{c}(t)$ is the
expectation value of $\hat{\sigma}_{z}$ conditioned on the whole
history of the detector outcome $X(t')$ for $t'\leq t$, \emph{$\delta X$}
is the maximum detector signal proportional to the measurement signal
amplitude $\sqrt{\overline{n}}$, and $\xi_{0}(t)$ is the total output
noise of the amplifier. Our test of inequality (\ref{eq:LG}) consists
in accurately measuring the steady state value of $K(\tau)=\langle(X(t)-\bar{X})(X(t+\tau)-\bar{X})\rangle_{t}/(\delta X/2)^{2}$
with a low measuring power, while the TLS is coherently driven.

\subsection*{Measurement-induced-dephasing in ensemble averaged Rabi oscillations}

Recent experiments have already investigated the back-action of the
measurement on a TLS with a similar circuit-QED setup \cite{Schuster}.
However only ensemble averaged quantities (i.e. obtained by averaging
the outcomes of many identical experimental sequences) had been considered
prior to this work, and the only detectable effect of a measurement
on the TLS dynamics was some extra dephasing, as demonstrated in \cite{Schuster}
by measuring the broadening of the TLS resonance line in presence
of a field in the resonator. This measurement back-action results
from the dependence of the TLS frequency $\omega_{ge}(t)=\omega_{ge}+2\chi n(t)$
on the photon number $n(t)$ stored in the resonator: fluctuations
of $n$ around $\bar{n}$ cause dephasing with a rate $\Gamma_{\phi}^{ph}(\bar{n})=8\bar{n}\chi^{2}/\kappa$
proportional to the measurement strength. We first perform a series
of control measurements to verify on ensemble-averaged Rabi oscillations
our quantitative understanding of the measurement-induced dephasing.
After a field of $\bar{n}$ photons (see supplementary information\ref{sub:Calibration-of-nbar})
is established inside the resonator using $V_{m}$, a Rabi pulse of
duration $\Delta t$ is applied to the TLS with $V_{d}$ (see Fig.~\ref{fig2}B),
followed by a strong measurement pulse. The phase $\langle\varphi(\Delta t)\rangle$
of the reflected measurement pulse is averaged over an ensemble of
typically $10^{4}$ identical experimental sequences, yielding the
data displayed in Fig.~\ref{fig2}B. One observes that for sufficiently
low measurement strength $\bar{n}$, the coherent dynamics is only
weakly affected by the measurement. This is the regime where the Bell's
inequality in time can be tested. For stronger measurement strength,
the oscillations are progressively washed out and replaced by an exponential
damping. For even stronger measurement strengths, the characteristic
time of the exponential becomes longer and longer (see Fig.~\ref{fig2}B),
revealing that a strong measurement inhibits the transition of the
TLS from ground to excited state as expected from the quantum Zeno
effect \cite{MisraSudarshan,Bernu}. We checked that these data are
in quantitative agreement with the expected measurement-induced dephasing
(see supplementary information\ref{sub:Measurement-induced-dephasing}).
However, it is important to realize that this set of measurements
would be unchanged if our driven quantum TLS was replaced by a precessing
classical spin, such as a macroscopic ferromagnet. Indeed, a macrospin
obeys similar equations of motion as the expectation value of the
spin of a TLS, namely Bloch equations. Thus no conclusion about the
correlations between measurements at different times can be drawn
that would allow a test of inequality (\ref{eq:LG}).

\subsection*{Continuous measurement of Rabi oscillations in the frequency domain}

We measure these correlations by monitoring the system in its steady
state, long after the transient ensemble averaged Rabi oscillations
such as shown in Fig.~2B have been washed out. Instead of applying
microwave pulses, the sources $V_{d}$ and $V_{m}$ are now continuously
ON. The quantity of interest is the two-time correlation function
$K(\tau)$, whose direct calculation from the measured time traces
$X(t)$ is difficult in our setup because the amplifier noise dominates
the output signal. However, this added noise can be removed by processing
the signal in the frequency domain.

\begin{figure}
\includegraphics[width=8.8cm]{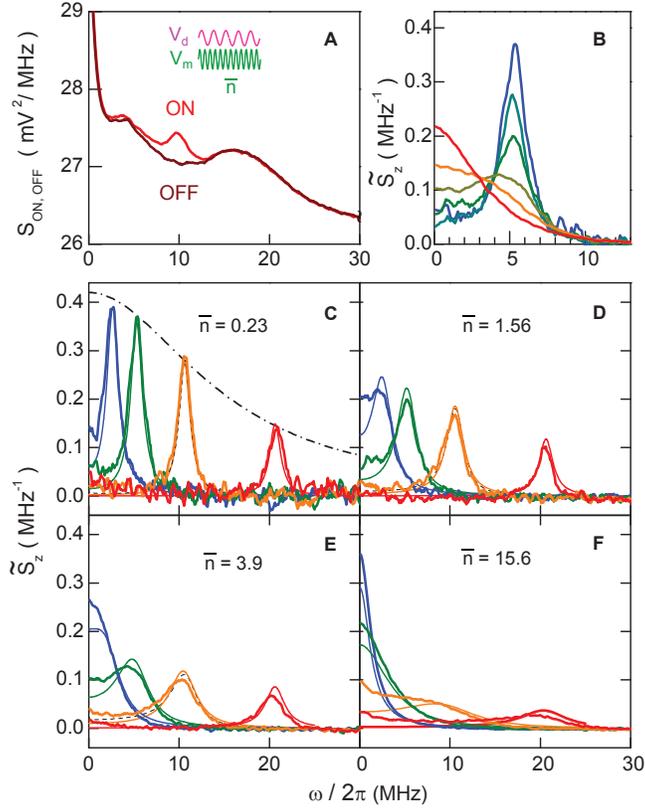}

\caption{Continuous monitoring of the TLS driven at the Rabi frequency $\omega_{R}$
for different measurement strengths $\bar{n}$. Each power spectrum
is acquired in $40$ to $80$ minutes. (A) Spectral densities $S_{ON}(\omega)$
and $S_{OFF}(\omega)$ when $V_{d}$ and $V_{m}$ are both OFF (brown)
or both ON (red); here $\omega_{R}/2\pi=10\,$MHz and $\bar{n}=1$.
The difference between ON and OFF shows a peak at the Rabi frequency.
(B) Normalized Rabi spectra $\tilde{S}_{z}(\omega)$ after correction
from the frequency response of the measuring line and conversion of
the output voltage into units of $\sigma_{z}$, at $\omega_{R}/2\pi=5\,$MHz
and $\bar{n}=0.23$, $0.78$, $1.56$, $3.9$, $7.8$, and $15.6$
(from blue to red). The curves show the weak to strong measurement
transition. (C-F) Normalized Rabi peaks at $\omega_{R}/2\pi=2.5$,
$5$, $10$ and $20\,$MHz (from blue to red) for $\bar{n}=0.23$
(C), $1.56$ (D), $3.9$ (E), and $15.6$ (F). Thick and thin color
lines are respectively the experimental spectra and those calculated
from a theoretical analytical formula (see text and supplementary
information\ref{sub:spectra-formula}) using only independently measured
parameters (including $\chi/2\pi=1.8\,$MHz). Dashed black lines on
top of the orange curves in (C,D,E) are Rabi peaks obtained by numerical
simulation with the same parameters. The dotted-dashed black curve
in C is the Lorentzian frequency response $C(\omega)$ of the resonator.}

\label{fig3} 
\end{figure}

For this purpose, we compute the square modulus $S_{I}$ and $S_{Q}$
of the Fourier transforms of $I(t)$ and $Q(t)$, to obtain the detector
output power spectrum $S(\omega)=S_{I}(\omega)+S_{Q}(\omega)$. The
signal power spectrum is then obtained by subtracting the amplifier
noise spectrum $S_{OFF}(\omega)$ measured when the two sources $V_{d}$
and $V_{m}$ are OFF from the signal-plus-noise spectrum $S_{ON}(\omega)$
measured with both sources ON, and by dividing this difference by
the independently measured frequency response $R(\omega)$ of the
measuring line (see supplementary information\ref{sub:line frequency response}).
Typical curves are shown in Fig.~\ref{fig3}A. They show a single
peak located at the Rabi frequency (already known from the time-domain
measurements), without any harmonics within the $50$~MHz detection
window. The output spectrum of a continuously monitored TLS undergoing
coherent oscillations \cite{Ilichev,Deblock,Manassen} has been the
subject of a number of theoretical calculations \cite{Korotkov_Averin,Goan,Shnirman},
and precise knowledge of all sample parameters allows us to obtain
for the first time a quantitative comparison with these theories.
Indeed, we can convert the signal power spectrum in spin units by
dividing it by a conversion factor $(\delta V/2)^{2}$ measured in
a calibration experiment by saturating the $g-e$ transition (see
supplementary information\ref{sub:Calibration-of-spectra}). The variation
of the resulting $\tilde{S}_{z}(\omega)$ spectrum with increasing
measurement power is shown in Fig.~\ref{fig3}B and is in good agreement
with theoretical predictions \cite{Korotkov_Averin,Goan,Shnirman}.
These data clearly show the transition from weak to strong measurement
in a continuously monitored driven TLS: at low $\bar{n}$, the spectrum
consists of a single Lorentzian peak at $\omega_{R}$; upon increasing
the measurement strength, the Lorentzian broadens towards low frequencies,
and for strong measurements, the spectrum becomes a Lorentzian centered
at zero frequency, similar to that of an incoherent TLS jumping stochastically
between its two states. In terms of quantum trajectories, the Lorentzian
spectra obtained in these two regimes are indirect signatures of the
weak measurement-induced quantum phase diffusion along the Rabi trajectory,
and of the quantum jumps made by the spin during the strong measurement.
The theoretical curves shown in Fig.~\ref{fig3} are obtained using
an analytical formula derived from the solution of Bloch equations
\cite{Torrey} in which the finite detector bandwidth is taken into
account phenomenologically (see supplementary information\ref{sub:spectra-formula});
the accuracy of this formula was checked by direct numerical integration
of the system's master equation (see Fig.~\ref{fig3}C-E and\ref{sub:Numerical-simulations}).
The agreement between theory and experiment is good for $\bar{n}\leq5$
but is only qualitative at larger $\bar{n}$, possibly due to a breakdown
of the dispersive approximation.

\subsection*{Experimental test of the Bell's inequality in time}

\begin{figure}[tbh]
\includegraphics[width=13.5cm]{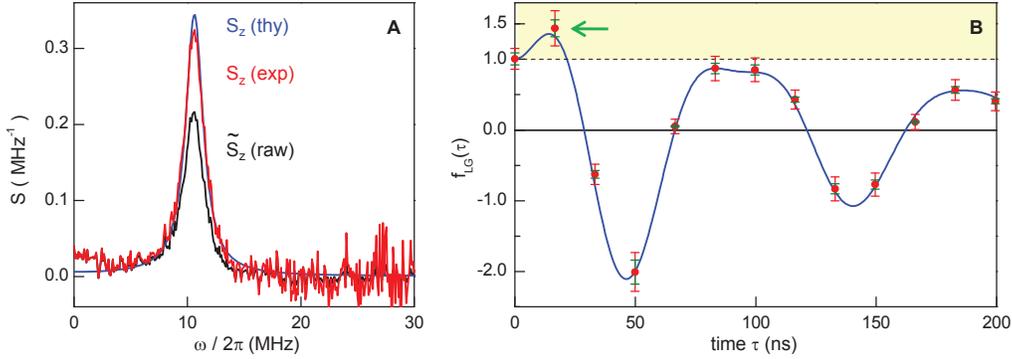}

\caption{Experimental violation of the {}``Bell's inequality in time'' introduced
in Fig.~\ref{fig1}B. A) experimental (red) and theoretical (blue)
spectral densities $S_{z}$, calculated or measured at $\omega/2\pi=10.6\,$MHz
and $\bar{n}=0.78$. The experimental curve is obtained by correcting
the raw $\tilde{S}_{z}$ spectrum ( black line) , acquired in 13 hours
with a $30\,$MHz bandwidth, from the frequency response $C(\omega)$
of the resonator (see Fig.~\ref{fig3}C). The blue curve is calculated
with $\Gamma_{1}^{-1}=200\,$ns and $\Gamma_{2}^{-1}=150\,$ns (see
supplementary information\ref{sub:spectra-formula}). B) experimental
(dots) and theoretical (blue line) Leggett-Garg quantity $f_{LG}(\tau)=2K(\tau)-K(2\tau)$,
with $K(\tau)$ the signal autocorrelation function obtained by inverse
Fourier transform of the $S_{z}$ curves in the left panel. Green
error bars correspond to the maximum systematic error associated with
calibration and $C(\omega)$, whereas red ones also include a two
standard deviation wide statistical error $\pm2\sigma(\tau)$ associated
with the experimental noise on $\tilde{S}_{z}$. The Leggett-Garg
inequality is violated (yellow region) at $\tau=17\,$ns (see green
arrow) by $5\sigma$.}

\label{Fig. 4} 
\end{figure}

We now turn to the test of inequality (\ref{eq:LG}). We measure a
Rabi peak at $\omega_{R}/2\pi=10.6\,$MHz with $\overline{n}=0.78$
photons and a $30\,$MHz detection window. Under macrorealistic assumptions,
the only effect of the bandwidth of the resonator would be to reduce
the measured signal by its Lorentzian response function $C(\omega)=1/[1+(2\omega/\kappa)^{2}]$;
we thus have to correct for this effect by dividing the measured spectral
density $\tilde{S}_{z}(\omega)$ by $C(\omega)$. We then compute
$K(\tau)$ by inverse Fourier transform of $S_{z}(\omega)=\tilde{S}_{z}(\omega)/C(\omega)$.
The experimental and theoretical Rabi peaks as well as the corresponding
$f_{LG}(\tau)$ curves are plotted in Fig.~\ref{Fig. 4}, showing
good overall agreement despite residual low-frequency noise possibly
originating from low-frequency fluctuations of $\omega_{ge}$. The
error bars on $f_{LG}(\tau)$ are the sum of the systematic errors
in the calibration of $\delta V$, $\kappa$, and $R(\omega)$, and
of the statistical error on the measured spectrum (see supplementary
information \ref{sub:Error-bars}).

We first note that we find $K(0)=f_{LG}(0)=1.01\pm0.15$, a value
close to 1 that directly results from the independent calibration
of $\delta V$. Since $K(0)$ represents the variance of $z(t)$ and
$|z(t)|\leq1$, this confirms that at any time $z(t)=\pm1$ as expected
for a quantum TLS. A classical macrospin oscillating as shown in Fig.\ \ref{fig3}
and calibrated with the same method would have given a variance of
1/2 instead. Note also that $K(0)=1$ could never be deduced from
ensemble averaged Rabi oscillations such as those of Fig.~\ref{fig2}B.
This is an example of the specific interest of correlation-function
measurements compared to time-domain ensemble averaged signals. Most
importantly, we observe that $f_{LG}(\tau)$ goes above the classical
limit of $1$, reaching $1.44(\pm0.12)\,\pm2\times(\sigma=0.065)$
at $\tau=17\,$ns$\,\thicksim\pi/3\omega_{R}$ and thus violating
inequality (\ref{eq:LG}) by 5 standard deviations $\sigma$. This
maximum of $f_{LG}$, slightly below the ideal value of $1.5$ in
absence of decoherence, is a direct signature of the invasive character
of the measurement process, which projects partially but continuously
the TLS towards the state corresponding to the detector output. It
is the interplay between this continuous projection and the coherent
dynamics that yields the violation of the inequality. More quantitatively,
the maximum of $f_{LG}$ is in agreement with the quantum prediction
of $1.36$ when taking into account the independently measured relaxation
and dephasing rates of the TLS. This violation of the Leggett-Garg
inequality rules out a simple interpretation of $K(\tau)$ as the
correlation function of a classical macrospin. It therefore brings
further evidence that a collective degree of freedom characterizing
a Josephson circuit can behave quantum-mechanically. It also demonstrates
that the back-action of a weak measurement, far from being a simple
noise that spoils quantum coherence as could be deduced from ensemble
averaged measurements, tends also to reinforce correlations between
measurements made at different times.

It is interesting to discuss in what respect the assumptions made
in analysing the experimental data influence the final result: apart
from simple corrections relying on classical electromagnetism, we
determine the main normalization factor $\delta V$ by saturating
the TLS transition and assuming that the ensemble averaged spin, either
classical or quantum, obeys Bloch equations. We checked this assumption
in Fig.~\ref{fig2} ~and supplementary Fig.~\ref{fig7}, which
show in particular that the excursion of the signal when driving the
spin is symmetric around the saturation value, as it would be for
classical macrospins. The observed violation is thus not an artefact
of our analysis framework, and represents more than a mere self-consistency
check of a quantum model.

\subsection*{Conclusion }

In conclusion, we have reported the experimental violation of a {}``Bell's
inequality in time'' by continuously monitoring the state of a superconducting
artificial TLS while it performed Rabi oscillations. The measured
two-time correlation function of the detector output reveals strong
non-classical correlations between the signal already recorded and
the TLS subsequent evolution. Our work thus brings a further proof
of the truly quantum-mechanical character of Josephson artificial
atoms. It is moreover a first step towards the test of a number of
important predictions for such a system \cite{Korotkov_Averin}, and
towards certain quantum feedback schemes: if the continuous monitoring
could be performed with a quantum-limited amplifier \cite{Qlim_amplifier1,Qlim_amplifier2},
it would become indeed possible to stabilize the phase of the Rabi
oscillations by feeding back the demodulated signal onto the amplitude
or frequency of the source that drives the TLS. This would modify
the shape of the Rabi peak, on top of which a narrow line should develop
\cite{feedback}. The correlations demonstrated in this work could
then constitute a key resource for quantum feedback, in the same way
as entanglement is a resource for quantum information processing.
\begin{acknowledgments}
We acknowledge financial support from European projects EuroSQIP and
SCOPE, and from ANR project Quantjo and C'Nano Ile-de-France for the
nanofabrication facility at SPEC. We gratefully thank P. Senat, P.
Orfila and J.-C. Tack for technical support, and acknowledge useful
discussions within the Quantronics group and with A. Lupascu, A. Wallraff
and R. Ruskov.

\bigskip{}

Author contributions: A.K., P.B., \& A.P.L. did the theoretical work,
A.P.L., F.M., P.B., D.V., \& D.E. designed the experiment, A.P.L.
fabricated the sample, A.P.L., F.M., P.B., \& F.N. performed the measurements,
A.P.L., F.M., D.V., \& P.B. analyzed the data, and all the authors
contributed to the writing of the manuscript.

\bigskip{}

Correspondance should be addressed to P.B.\end{acknowledgments}

\newpage

\section*{Supplementary Information}

\subsection{Sample fabrication, setup, and measurement protocol\label{sub:Sample-fabrication}}

The sample is fabricated using standard lithography techniques. In
a first step, a $200\,$nm Niobium thin-film is sputtered on a high-resistivity
oxidized Silicon substrate. It is patterned with optical lithography
followed by reactive ion etching of the Niobium to form the Coplanar
Waveguide resonator. The transmon is then patterned by e-beam lithography
followed by double-angle evaporation of two Aluminum thin-films, the
first one being oxidized to form the junction oxide. The sample is
glued on a microwave printed-circuit board, enclosed in a copper box,
and thermally anchored to the mixing chamber of a dilution refridgerator
at typically $20\,$mK.

Measurement signals are generated by mixing the output of a microwave
source ($V_{d}$ or $V_{m}$) with DC pulses generated by arbitrary
waveform generators, using DC coupled mixers (not shown in Fig.~1).
They are then sent to the input microwave line that includes bandpass
filters and attenuators at various temperatures ($77\,$dB in total).
The output line contains a $4-8\,$GHz bandpass filter, a circulator
and two isolators (not shown in Fig.~1), and a cryogenic amplifier
CITCRYO1-12 (from Caltech) with $38\,$dB gain and noise temperature
$T_{N}=4\,$K. The output signals are further amplified at room-temperature
yielding a total gain of $56\,$dB, and finally mixed down using an
I/Q mixer with a synchronized local oscillator at the same frequency.
The $I$ and $Q$ quadratures are filtered with a $50\,$MHz low-pass
filter and further amplified with power gain $100$ ; the total gains
on channels $I$ and $Q$ are equated with a precision better than
0.5\%. Both quandratures are then sampled by a fast digitizer and
transferred to a computer that processes them.

In the case of ensemble-averaged Rabi oscillations, $200\,$ns long
measurement pulses are used; the corresponding $X(t)$ traces are
averaged over $10^{4}$ identical sequences (repetition rate: $200\,$kHz),
yielding time traces as shown in Fig.~\ref{fig2}B.

In the case of power spectrum measurements, we compute the Fast Fourier
Transform $X(\omega)$ on records of $1024$ X(t) samples separated
by $10\,$ns. We average $S_{X}=|X(\omega)|^{2}$ over $10^{5}$-$10^{6}$
identical sequences, yielding spectra as shown in Fig.~3A. In these
experiments, it is crucial to accurately subtract the amplifier noise,
whose power is about $60$ times stronger than the Rabi peak we want
to measure. For that purpose, we measure separately and subtract the
signal+noise $S_{X\, ON}$ and the noise $S_{X\, OFF}$ by alternating
periods of duration $T_{ON}=T_{OFF}=2.5\,$ms, during which both microwave
sources $V_{d}$ and $V_{m}$ are ON, or both are OFF. Each time the
pulses are switched ON or OFF, we wait a time $T_{ss}=5\mu$s to let
the system reach its new steady state. The resulting experimental
sequence is shown in supplementary Fig.~\ref{fig5}. Note that this
subtraction also suppresses the influence of long-term drifts of the
amplifiers gains. 

\begin{figure}[t]
\includegraphics[width=8.59cm]{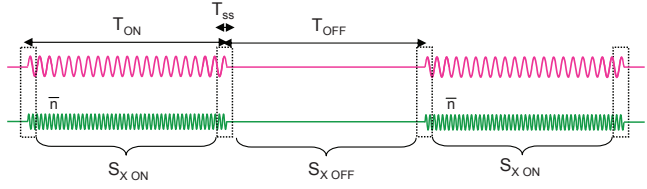}\\

\caption{Experimental procedure for measuring the signal power spectra. Both
microwave sources $V_{d}$ and $V_{m}$ are switched ON and OFF during
$T_{ON}=T_{OFF}=2.5\,$ms. Both quadratures are sampled, their Fast
Fourier Transform calculated and squared to compute the power spectra
$S_{X\, ON}$ and $S_{X\, OFF}$. Dashed rectangles show the $T_{SS}=5\mu$s
waiting times for establishing the system steady state.}

\label{fig5} 
\end{figure}

\subsection{Sample parameters determination\label{sub:Sample-parameters}}

The resonator parameters $\omega_{c}$ and $\kappa$ given in the
text, as well as the TLS frequency $\omega_{ge}$ are determined by
standard spectroscopic measurements and are consistent with the design
values. In all the experiments with no field in the resonator, the
TLS frequency is fixed at $\omega_{ge}/2\pi=5.304\,$GHz by tuning
the magnetic field. The $\chi$ value, which is used to calibrate
the mean photon number $\overline{n}$, is determined as follows:

When the TLS state changes from $g$ to $e$, the phase of the reflected
microwave signal at the resonator frequency varies by $2\delta\varphi_{0}=4\arctan(2\chi/\kappa)$.
We thus measure $\delta\varphi_{0}$ by applying a strong and long
driving pulse $V_{d}$ that saturates the $g$-$e$ transition and
results in an almost equal $50\%$ population of $g$ and $e$. In
this experiment, the resonator is probed continuously with a low-amplitude
field ($\overline{n}\sim2)$ while the reflected signal is measured
by heterodyne detection with a local oscillator detuned from $V_{m}$
by $3.2$MHz. The resulting beating pattern in the two quadratures
$X(t)$ is ensemble averaged over a few $10^{5}$ identical sequences,
and its phase $\varphi(t)$ is fitted and plotted in supplementary
Fig.~\ref{fig6}. The plot yields a shift $\delta\phi_{0}=12.5\pm0.2\lyxmathsym{\textdegree}$
between no driving and saturation.

\begin{figure}
\includegraphics[width=7cm]{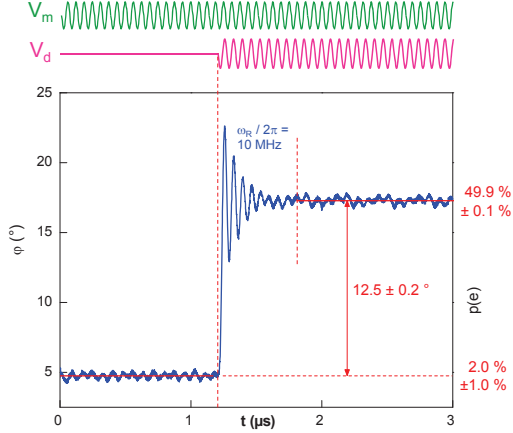}\\

\caption{Ensemble averaged measurement of $\varphi(t)$ when a driving pulse
saturates the $g-e$ transition starting from $g$. The measurement
is continuous with $\overline{n}\sim2$ photons.}

\label{fig6} 
\end{figure}

For converting this shift into a $\chi$ value, we take into account
two effects:

(i) the TLS is not perfectly in its ground state $g$ in the absence
of microwave drive due to residual thermal excitation to $e$. We
have measured this thermal population $p(e)_{0}=0.02\pm0.01$ by measuring
the noise spectrum with $V_{d}$ being switched OFF. 

(ii) the saturation induced by the driving field is slightly below
$50\%$ due to longitudinal and transversal relaxation. The population
$p(e)_{st}$ in the excited state at saturation is indeed given by
the steady state solution to Bloch equations

\begin{equation}
p(e)_{st}=\frac{1}{2}-[\frac{1}{2}-p(e)_{0}]\frac{1+(\Gamma_{2}^{-1}\delta\omega)^{2}}{1+(\Gamma_{2}^{-1}\delta\omega)^{2}+(\omega_{R}^{2}\Gamma_{1}^{-1}\Gamma_{2}^{-1})}.\end{equation}
Here $\omega_{R}/2\pi=10$ MHz is fixed by the driving strength, $\Gamma_{1}^{-1}=200\pm10$
ns and $\Gamma_{2}^{-1}=150\pm10\,$ns are independantly measured,
and $\delta\omega$ is the residual detuning of the driving source
from the TLS resonance. In our experiment the driving frequency was
scanned to experimentally maximize $\delta\phi_{0}$ so that $\delta\omega\sim0$
. This yields $p(e)_{st}=0.496\pm0.001$ instead of $0.5$.

Taking (i) and (ii) into account, we thus calculate $\delta\phi_{0}=(0.95\pm0.02)\delta\varphi_{0}$.
Maximizing all the uncertainties including those on $\kappa$, we
finally obtain $\chi/2\pi=\kappa\tan(\delta\varphi_{0}/2)/4\pi=1.75(-0.11/+0.14)\,$MHz,
also consistent with the TLS parameters determined by spectroscopy.

\subsection{Calibration of the mean photon number $\overline{n}$ at frequency
$\omega_{c}$\label{sub:Calibration-of-nbar}}

The calibration of $\overline{n}$ as a function of the applied input
power $P$ of $V_{m}$ is needed for checking that the measurement-induced
dephasing is quantitatively understood (see D) and for producing the
theoretical curves of Figs.~\ref{fig3} and \ref{Fig. 4} without
any fitting parameters. To perform this calibration, we measure the
AC Stark shift$-2\chi\overline{n}$ of the TLS frequency $\omega_{ge}$
as a function of the input power $P$; $\chi$ being known (see\ref{sub:Sample-parameters}),
the shift provides an in-situ calibration of $\overline{n}(P)$.

Note that all the curves presented in this article were obtained with
the TLS driving source $V_{d}$ tuned in resonance with the AC Stark
shifted TLS frequency. For that, the TLS spectroscopy peak in presence
of the very same $\bar{n}$ field was always measured just before
each acquisition.

\subsection{Measurement-induced dephasing and Quantum Zeno Effect\label{sub:Measurement-induced-dephasing}}

\begin{figure}
\includegraphics[width=10cm]{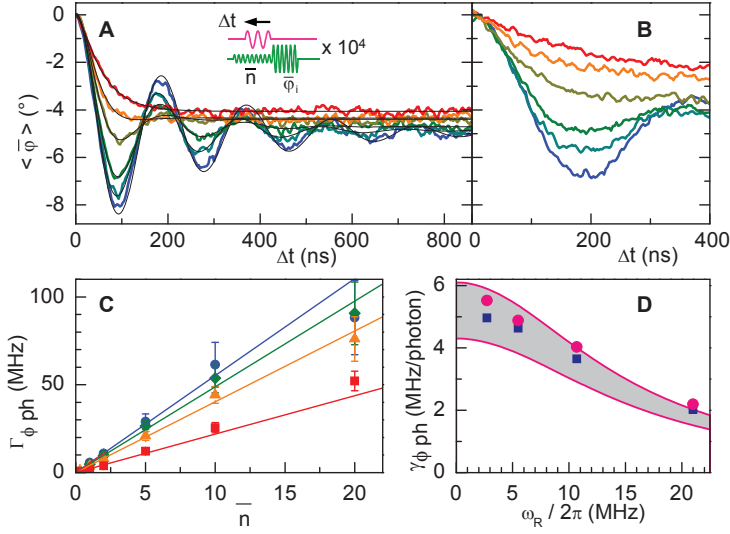}\\

\caption{Determination of the measurement-induced dephasing using ensemble-averaged
Rabi oscillations recorded with the protocol of Fig.~\ref{fig2}B.
(A) Same data as in Fig. \ref{fig2}B with $\omega_{R}/2\pi=5\,$MHz
and $\bar{n}=0$, $1$, $2$, $5$, $10$ and $20$ (blue to red).
Fits to the solution of Bloch equations (thin black lines) yield the
measurement-induced dephasing rate $\Gamma_{\phi}^{ph}$, after subtraction
of other decoherence contributions deduced from the curve at $\bar{n}=0$.
(B) Similar experimental data at $\omega_{R}/2\pi=2.5\,$MHz and same
photon numbers. When the measurement strength is increased, the time
dependence at short times changes from quadratic to approximately
linear with an increasing time constant, which is a manifestation
of the Quantum Zeno effect. (C) Experimental values (dots with error
bars) of $\Gamma_{\phi}^{ph}(\bar{n})$ for $\omega_{R}/2\pi=2.5$,
$5$, $10$ and $20\,$ MHz (blue to red). Solid lines are linear
fits of the data taken at $\bar{n}\leq5$ , yielding the measurement-induced
dephasing rate per photon $\gamma_{\phi}^{ph}$. (D) $\gamma_{\phi}^{ph}$
as a function of $\omega_{R}$. Comparison between the experimental
values (magenta dots), values obtained by numerical integration of
the system master equation (blue squares), and theoretical curves
(magenta lines) given by Eq. \ref{eq:GammaR} and using only measured
parameters (the two lines limiting the grey area correspond to the
lower and higher bounds of experimental uncertainties).}

\label{fig7} 
\end{figure}

Since measurement-induced dephasing is a key ingredient of the present
work, we checked our quantitative understanding of it. For that, we
measure ensemble-averaged Rabi oscillations (see supplementary Fig.~\ref{fig7})
in presence of a perturbing field of $\overline{n}$ photons at frequency
$\omega_{c}$, which mimicks the effect of a continuous measuring
field. In this experiment, the AC stark shifted TLS frequency $\omega_{ge}(t)=\omega_{ge}-2\chi n(t)$
fluctuates with the number of photons $n(t),$ which fluctuates because
of the shot noise in the measurement signal. In an ensemble-averaged
viewpoint, this leads to pure dephasing of the TLS with a dephasing
rate $\Gamma_{\phi}^{ph}(\bar{n})=8\bar{n}\chi^{2}/\kappa$ already
observed in other experiments \cite{Schuster}. Now, because Rabi
oscillations at $\omega_{R}$ are most sensitive to the noise spectral
density at $\omega_{R}$ \cite{Ithier} and because the shot noise
in the measurement signal is filtered by the resonator response $C(\omega)=1/[1+(2\omega/\kappa)^{2}]$,
one expects for the Rabi oscillations a pure dephasing rate

\begin{equation}
\Gamma_{\phi}^{ph}(\omega_{R},\bar{n})=\Gamma_{\phi}^{ph}(\bar{n})C(\omega_{R}).\label{eq:GammaR}\end{equation}

Supplementary Fig.~\ref{fig7} confirms all these expectations: the
Rabi oscillations are progressively washed out and replaced by an
exponential damping when the measurement strength $\overline{n}$
is increased. For even stronger measurements (larger $\overline{n}$),
the damping time constant increases (see Fig.~\ref{fig7}B), indicating
an inhibition of the TLS transition from $g$ to $e$ that is a signature
of the Quantum Zeno Effect. By fitting each Rabi curve with the analytical
solution of Bloch equations \cite{Torrey} (using the measured $\Gamma_{1}^{-1}=225\pm10\,$ns),
we obtain in Fig.~\ref{fig7}C the total decoherence rates $\Gamma_{2}(\omega_{R},\bar{n})$
that includes the measurement-induced dephasing $\Gamma_{\phi}^{ph}(\omega_{R},\bar{n})$
and the contribution $\Gamma_{2}(\omega_{R},0)=\Gamma_{\phi}^{0}(\omega_{R})+\Gamma_{1}/2$
from other dephasing sources and from energy relaxation ($\Gamma_{\phi}^{0}(2\pi\times5\,\mathrm{MHz})=(810\pm20\,\mathrm{ns})^{-1}$
in Fig. \ref{fig7}A for instance). For each $\omega_{R}$, the measured
$\Gamma_{\phi}^{ph}(\omega_{R},\bar{n})$ are proportional to $\bar{n}$
as expected. Their slopes $\gamma_{\phi}^{ph}(\omega_{R})$ determined
by fitting $\Gamma_{\phi}^{ph}(\omega_{R},\bar{n})$ up to $\bar{n}=5$
are shown in Fig.~\ref{fig7}D and are in good agreement with the
predictions of Eq. \ref{eq:GammaR} using the independendly measured
values of $\chi$ and $\kappa$. We thus have a quantitative understanding
of the measurement-induced dephasing in our system.

\subsection{Determination of the frequency response $R(\omega)$ of the measuring
line \label{sub:line frequency response}}

The measuring line between the sample and the output signal includes
several microwave circulators, filters, amplifiers, an IQ demodulator,
two fast digitizers, and many sections of cables with connectors.
Its frequency response $R(\omega)$ was measured in-situ in order
to correct precisely the raw measured spectra $S(\omega)$. It is
shown in supplementary Fig.~\ref{frequency response of the line}
after normalization to 1 at zero frequency. The error bar shown ($\pm1.5\%$)
is an upper bound of the maximum systematic error over the whole frequency
range.

\begin{figure}[h]
\includegraphics[width=8.59cm]{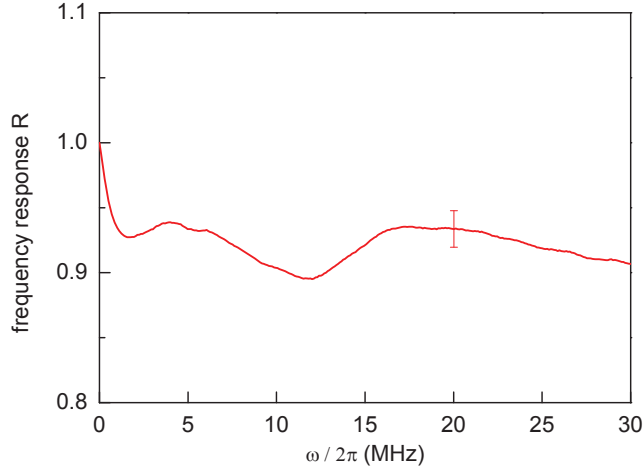}\\

\caption{Frequency response $R(\omega)$ of the measuring line, including the
amplification and demodulation chain. The error bar represent a constant
maximum relative error.}

\label{frequency response of the line} 
\end{figure}

\subsection{Calibration of the spectra : $\delta V(\overline{n})$ \label{sub:Calibration-of-spectra}}

\begin{figure}
\includegraphics[width=8.59cm]{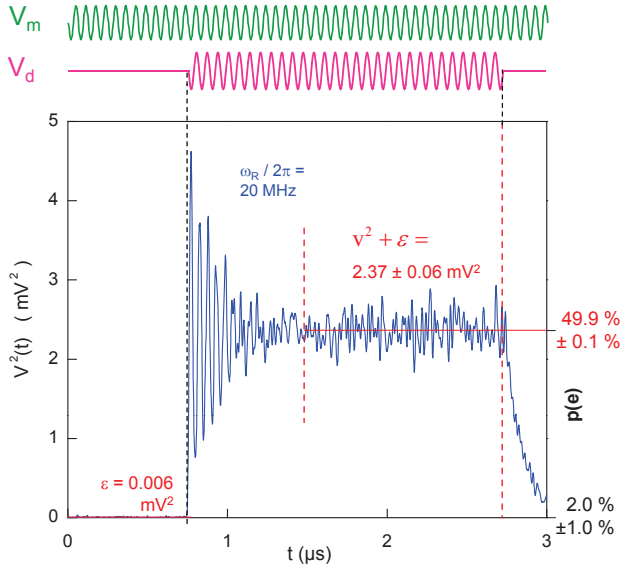}\\

\caption{Calibration of the conversion factor $\delta V/2$ between output
demodulated voltage and spin units. A measurement pulse $V_{m}$ is
applied with an amplitude corresponding to $\bar{n}=0.78$. After
about one \textmu{}s, a saturating pulse is applied at $\omega_{ge}$
during about 2 \textmu{}s. The $V^{2}(t)$ signal is measured and
averaged over a few $10^{5}$ identical sequences. Starting from a
thermal mixture of 98\% $g$ and 2\% $e$, the TLS undergoes Rabi
oscillations at about 20 MHz before it reaches its steady state with
50\% $g$ and $e$ population. The steady-state output yield $\delta V$
(see text).}

\label{fig. Calibration deltaV} 
\end{figure}

According to the definition of $S(\omega)$, the conversion factor
$\delta V(\overline{n})/2$ between the demodulated output signals
in Volts and spin units is defined as $[\delta V(\overline{n})]^{2}=[\delta I(\overline{n})]^{2}+[\delta Q(\overline{n})]^{2}$,
with $\delta X(\overline{n})$ the change in quadrature $X$ when
the TLS state changes from $g$ to $e$. This definition has the great
advantage of being insensitive to any drift or jitter of the relative
microwave phase between the measurement source $V_{m}$ and the local
oscillator used for the demodulation. Since $(\delta V/2)^{2}$ is
the normalization factor of the spectrum used to test inequality (4),
it is particularly important to determine it with the best possible
precision. Although it could be calculated from several independently
measured parameters, we found more accurate to calibrate it by direct
measurement: we thus ensemble average $V^{2}(t)=[I_{ON}(t)-I_{OFF}(t)]^{2}+[Q_{ON}(t)-Q_{OFF}(t)]^{2}$
under saturation of the $g$-$e$ transition (see supplementary Fig.~\ref{fig. Calibration deltaV}),
similarly to what is reported in B to determine $\chi$. In this experiment,
$V_{m}$ is always ON and only $V_{d}$ is switched ON and OFF, and
$\omega_{R}/2\pi=20$ MHz so that $p(e)_{st}=0.499\pm0.001$; $V^{2}(t)$
varies from $\epsilon=4<\xi_{0}^{2}>$ (with $<\xi_{0}^{2}>$ the
variance of the noise on each $X$ up to $v^{2}+\varepsilon$ with
$v=[p(e)_{st}-p(e)_{0}]\delta V$ when saturation is reached. From
supplementary Fig.~\ref{fig. Calibration deltaV}, we obtain $(\delta V)^{2}=10.29\pm0.64\,\mathrm{mV^{2}}$
for $\overline{n}=0.78$. It is important to note here that this calibration
was done with the very same measurement power as that used to record
the spectrum of Fig.~\ref{Fig. 4}, so that the uncertainty on $\overline{n}$
does not impact at all the violation of the Leggett-Garg inequality.
Moreover, as averaging of this particular spectrum took about 13 hours,
the calibration was performed twice, before and after averaging, in
order to check that the experiment was stable: the second calibration
yields $(\delta V)^{2}=10.59\pm0.64\,\mathrm{mV^{2}}$. We thus take
$(\delta V/2)^{2}=2.61\pm0.16\,\mathrm{mV^{2}}$ for the conversion
factor in our test of inequality (\ref{eq:LG}).

The experimental spectra of Fig.~3 obtained for different $\bar{n}$
have also been expressed in spin units by simply rescaling $\delta V(\bar{n}=0.78)$
by $\sqrt{\overline{n}/0.78}$ at low $\bar{n}$. For $\bar{n}$ above
3, we also take into account corrections to the dispersive approximation,
which is valid only for $\bar{n}\ll n_{crit}=\Delta^{2}/4g_{0}^{2}=31$
\cite{Blais} in our case (where $g_{0}$ is the TLS-resonator coupling
constant). Using a model similar to \cite{Gambetta}, we keep the
form of the dispersive Hamiltonian unchanged but use a modified dispersive
constant $\chi(\bar{n})=\chi(0)(1-\lambda\bar{n})$ yielding a modified
conversion factor $\delta V(\bar{n})(1-\lambda\bar{n})$. We determine
$\lambda=7\cdot10^{-3}$ both experimentally and theoretically. Note
that this correction gives noticeable effect only for $\bar{n}=15$
in Fig.~\ref{fig3}F.

\subsection{Analytical formula for the frequency spectra\label{sub:spectra-formula}}

The noise spectrum of a TLS undergoing coherent oscillations under
continuous measurement has been computed in the case of a quantum
dot coupled to a quantum point contact. Such a problem can be completely
mapped onto our experiment with a transmon in a resonator with infinite
bandwidth. In this limit, an exact analytical formula exists for the
spectrum $S_{\hat{z}}(\omega)$ \cite{Korotkov_Averin}, which is
simply the Fourier transform of the two-time correlation function
$K_{\hat{z}}(\tau)=\langle\hat{\sigma_{z}}(t)\hat{\sigma_{z}}(t+\tau)\rangle_{t}$
obtained from the analytical solutions to the Bloch equations:

\begin{multline}
S_{\hat{z}}(\omega)=\frac{4}{[\gamma^{2}+(\omega-\tilde{\omega}_{R})^{2}][\gamma^{2}+(\omega+\tilde{\omega}_{R})^{2}]}\cdot\Bigr\{\gamma(1-z_{st}^{2})(\gamma^{2}+\tilde{\omega}_{R}^{2}+\omega^{2})+\\
+\left[(1-z_{st}^{2})(\Gamma_{2}-\Gamma_{1})/2-\omega_{R}^{2}z_{st}^{2}/\Gamma_{2}\right](\gamma^{2}+\tilde{\omega}_{R}^{2}-\omega^{2})\Bigr\},\label{eq:sz}\end{multline}

\noindent where $\tilde{\omega}_{R}=\sqrt{\omega_{R}^{2}-(\Gamma_{2}-\Gamma_{1})^{2}/4}$,
$\gamma=(\Gamma_{2}+\Gamma_{1})/2$, and $z_{st}=-1/(1+\Gamma_{1}^{-1}\Gamma_{2}^{-1}\omega_{R}^{2})$
is the steady state solution. Due to the finite bandwidth of the detector,
the measured signal $\tilde{S}_{z}$ is reduced as well as the dephasing
rate according to Eq. (\ref{eq:GammaR}). We modify Eq. \ref{eq:sz}
to take this into account by (i) multiplying $S_{\hat{z}}$ by the
Lorentzian cutoff $C(\omega)=1/(1+(2\omega/\kappa)^{2})$ and (ii)
by changing the dephasing contribution $\Gamma_{\phi}$ to the total
decoherence rate $\Gamma_{2}=\Gamma_{\phi}+\Gamma_{1}/2$ into $\Gamma_{\phi}C(\omega)$;
we obtain in this way an expression $\tilde{S}_{\hat{z}}(\omega)$.
The validity of this phenomenological approach was checked by numerical
simulations as explained in \ref{sub:Numerical-simulations}. The
calculated spectra of Fig.~\ref{fig3}C-F and in the inset of Fig.~\ref{Fig. 4}
are obtained using the independently measured values $\Gamma_{1}=(200\,\mathrm{ns)^{-1}}$and
$\Gamma_{\phi}C(\omega)=(150\,\mathrm{ns)^{-1}}$ (see \ref{sub:Measurement-induced-dephasing}).

\subsection{Numerical simulations\label{sub:Numerical-simulations}}

\begin{figure}[h]
\includegraphics[width=8.59cm]{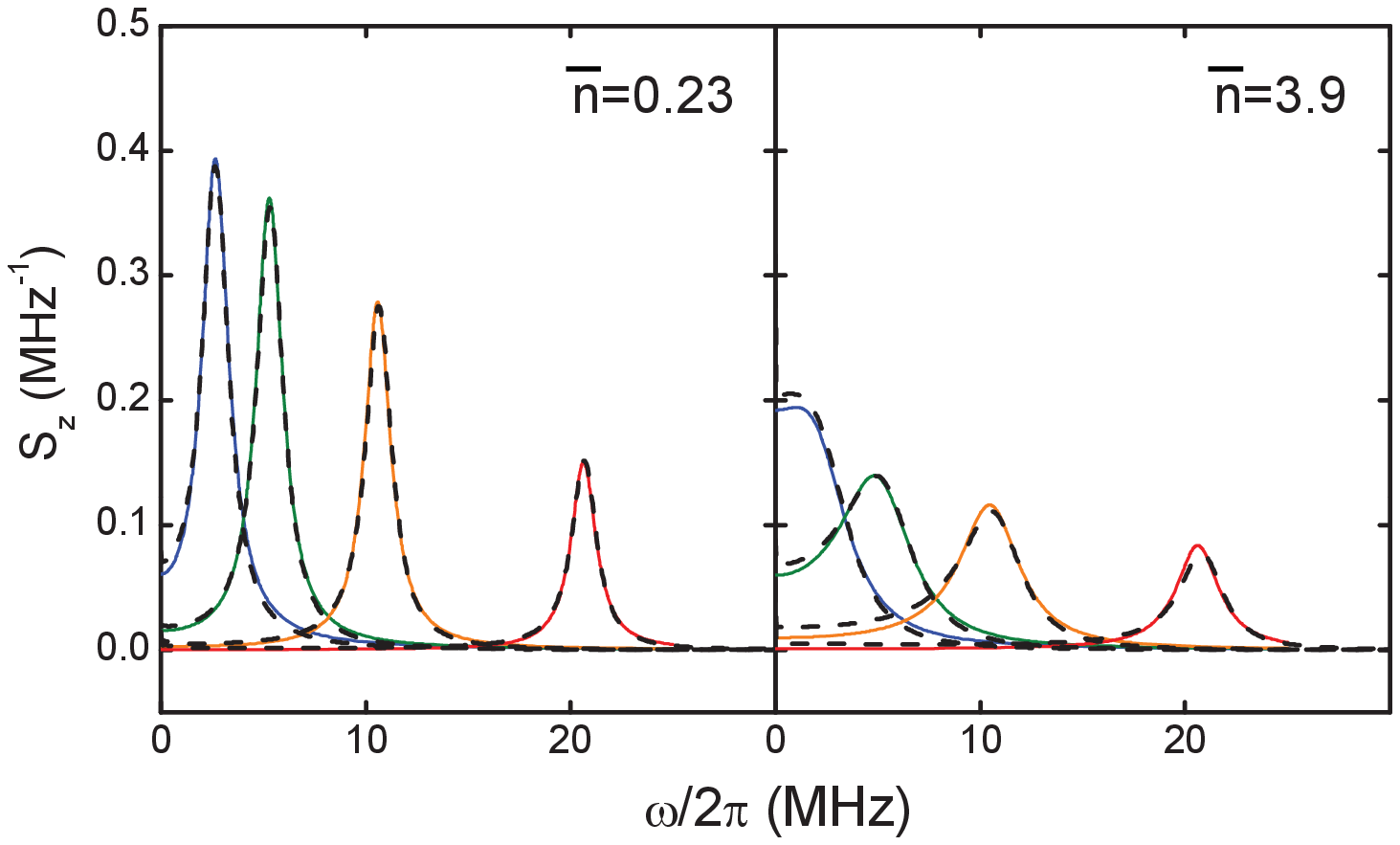}\\

\caption{Comparison between the analytical spectra $\tilde{S}_{\hat{z}}$ (solid
lines) and numerical simulations using the master equation as explained
in the text (dashed black lines), for $\bar{n}=0.23$ and $3.9$,
and $\omega_{R}/2\pi=2.5,5,10,20\,$MHz. The agreeement is excellent.}

\label{fig10} 
\end{figure}

We perform numerical calculations of the combined TLS-cavity field
evolution by integrating the master equation of the system (as described
in \cite{Gambetta}). We model its coherent evolution with the effective
dispersive Hamiltonian (valid in the dispersive limit $|\Delta|\gg g_{0}$)

\begin{equation}
H_{eff}/\hbar=-\frac{(\omega_{ge}-\omega_{d})}{2}\sigma_{z}+(\omega_{c}-\omega_{m})a^{\dag}a+\chi a^{\dag}a\sigma_{z}+V_{m}(a^{\dag}+a)+V_{d}(\sigma_{-}+\sigma_{+})\end{equation}

\noindent with $\omega_{d}$ and $\omega_{m}$ the frequencies of
the sources $V_{d}$ and $V_{m}$ driving the TLS and the resonator.
Taking into account the damping of the resonator field at rate $\kappa$
and of the TLS energy and coherence at rates $\Gamma_{1}$ and $\Gamma_{\phi}$,
the master equation is

\begin{equation}
\dot{\rho}=-\frac{i}{\hbar}[H_{eff},\rho(t)]+\kappa\mathcal{D}[a]\rho(t)+\Gamma_{1}\mathcal{D}[\sigma_{-}]\rho(t)+\Gamma_{\phi}\mathcal{D}[\sigma_{z}]\rho(t)/2,\end{equation}

\noindent with $\mathcal{D}[A]\rho=A\rho A^{\dag}-A^{\dag}A\rho/2-\rho A^{\dag}A/2$.
We integrated this equation using a quantum optics library available
online \cite{qotoolbox}.

We used this numerical tool to perform simulations of Rabi oscillations
dephased by $\bar{n}$ photons in the cavity (dark blue squares in
supplementary Fig.~\ref{fig7}d), taking $\chi/2\pi=1.8\,$MHz. We
also used it to compute the two-time correlation function of the detector
output, i.e. $K'(\tau)=\kappa\langle Re[a^{\dag}(t+\tau)a(t)]\rangle_{t}$,
by using the quantum regression theorem \cite{MilburnWalls}; we then
convert $K'(\tau)$ into a power spectrum by Fast Fourier Transform.
The conversion into spin units was done as in the experiment, by calculating
the output signal at saturation $\delta V$ of a continuously monitored
Rabi oscillation. We were able to perform the simulations only up
to $\bar{n}=7.8$, yielding the dashed lines shown in Fig.~\ref{fig3}C-F.
The agreement with the analytical formula is excellent for all calculated
curves, as can be seen in Fig.~\ref{fig10}.

\subsection{Experimental uncertainties in the determination of $f_{LG}(\tau)$\label{sub:Error-bars}}

As mentionned in the text, the experimental points $f_{LG}(\tau)=2K(\tau)-K(2\tau)$
of Fig.~\ref{Fig. 4} are obtained from the inverse Fourier transform
$K(\tau)$ of $S_{z}(\omega)=\tilde{S}_{z}(\omega)/C(\omega)$ with 

\[
\tilde{S}_{z}(\omega)=\frac{S_{ON}(\omega)-S_{OFF}(\omega)}{R(\omega)(\delta V/2)^{2}}\,,\]
with $R(0)=C(0)=1$ and $\delta V/2$ being measured at zero frequency.
The systematic error bars on $f_{LG}(\tau)$ (in green in Fig.~\ref{Fig. 4})
result from the sum of the three maximum relative uncertainties, $\Delta R/R=\pm1.5\%$
, $\Delta(\delta V/2)^{2}/(\delta V/2)^{2}=\pm6.1\%$ and $\Delta C/C=2(\Delta\kappa/\kappa)/[1+(\kappa/2\omega)^{2}]$
with $\Delta\kappa/\kappa=\pm2.6\%$. Note that the frequency dependent
error $\Delta C/C$ is propagated exactly through the calculation
of $f_{LG}(\tau)$ and contributes to $\Delta f_{LG}/f_{LG}$ by $\pm0.8\%$
where inequation (\ref{eq:LG}) is violated. The total systematic
error at that point is thus $\pm8.4\%$. The statistical standard
deviation on each $f_{LG}(\tau)$ data point is computed by propagating
the statistical error on the measured Rabi spectrum (i) through the
division by the cavity filtering $C(\omega)$, (ii) through the definition
of the inverse Fourier transform, and (iii) through the difference
$2K(\tau)-K(2\tau)$. Each point $k=1$ to $N$ of the $S_{z}(\omega)$
spectrum with bin size $\Delta f=100\,$kHz has a constant standard
deviation $\sigma_{0}$ measured in the $22-30$ MHz region where
the spectral density is zero. Consequently, the standard deviation
on each point $k$ of the corrected spectrum is $\sigma_{k}=\sigma_{0}/C[2\pi\Delta f(k-1)]$.
Finally, the standard deviation $\sigma_{r}$ on each point $r$ of
$f_{LG}[\tau=(r-1)/(N\Delta f)]$ is

\begin{equation}
\sigma_{r}=\Delta f\sqrt{\sigma_{k=1}^{2}+4\sum_{k=2}^{N/2}\sigma_{k}^{2}\left[2\cos\frac{2\pi(r-1)(k-1)}{N}-\cos\frac{2\pi2(r-1)(k-1)}{N}\right]^{2}}.\end{equation}
A conventional $2\sigma_{r}$ statistical contribution is added to
the systematic error to form the total red error bars of Fig.~\ref{Fig. 4}.
At the second point $\tau=17ns$ where inequation (\ref{eq:LG}) is
violated, the standard deviation $s\equiv\sigma_{2}=0.065$, and the
bottom of the systematic error bar is $4.9\, s$ above 1.


\begin{thebibliography}{S3}
\bibitem{Bell}Bell, J.S., On the Einstein Podolvski Rosen Paradox.
\textit{Physics} (N.Y.) \textbf{1}, 195-200 (1965).

\bibitem{CHSH}Clauser, J.F., Horne, M.A., Shimony, A., \& Holt, R.A.
Proposed Experiment to Test Local Hidden-Variable Theories.\textit{
Phys. Rev. Lett.} \textbf{23}, 880-884 (1969).

\bibitem{Aspect}Aspect, A., Grangier, P., \& Roger, G. Experimental
Realization of Einstein-Podolsky-Rosen-Bohm Gedankenexperiment: A
New Violation of Bell's Inequalities. \textit{Phys. Rev. Lett.} \textbf{49},
91-94 (1982).

\bibitem{Chuang} Nielsen, M.A. \& Chuang, I.L. \textit{Quantum Computation
and Quantum Information}, Cambridge University Press (2000).

\bibitem{GargLeggett}Leggett, A.J. \& Garg, A. Quantum mechanics
versus macroscopic realism: Is the flux there when nobody looks? \textit{Phys.
Rev. Lett.} \textbf{54}, 857-860 (1985).

\bibitem{Ruskov}Ruskov, R., Korotkov, A.N., \& Mizel, A. Signatures
of Quantum Behavior in Single-Qubit Weak Measurements. \textit{Phys.
Rev. Lett.} \textbf{96}, 200404 (2006).

\bibitem{Goggin}Goggin, M. E. \textit{et al. }Violation of the Leggett-Garg
inequality with weak measurement of photons. arXiv:0907.1679.

\bibitem{Xu}Xu, J.S., Li, C.F., Zou, X.B., \& Guo, G.C. Experimentally
identifying the transition from quantum to classical with Leggett-Garg
inequalities, arXiv:0907.0176.

\bibitem{Korotkov_Averin} Korotkov, A.N. \& Averin, D.V. Continuous
weak measurement of quantum coherent oscillations. \textit{Phys. Rev.}
B \textbf{64}, 165310 (2001).

\bibitem{feedback}Ruskov R. \& Korotkov A. N. Quantum feedback control
of a solid-state qubit. \textit{Phys. Rev.} B \textbf{66}, 041401
(2002).

\bibitem{Martinis}Ansmann, M.\textit{ et al. }Violation of Bell's
inequality in Josephson phase qubits. \textit{Nature }\textbf{461},
504-506 (2009).

\bibitem{Rob Bell}Chow, J. M.\textit{ et al. }Entanglement Metrology
Using a Joint Readout of Superconducting Qubits. arXiv:0908.1955.

\bibitem{Blais}Blais, A., Huang, R., Wallraff, A., Girvin, S. M.,
\& Schoelkopf, R. J. Cavity quantum electrodynamics for superconducting
electrical circuits: An architecture for quantum computation. \textit{Phys.
Rev.} A \textbf{69}, 062320 (2004).

\bibitem{Wallraff}Wallraff, A. \textit{et al. }Strong coupling of
a single photon to a superconducting qubit using circuit quantum electrodynamics.
\textit{Nature} \textbf{431,} 162-167 (2004).

\bibitem{Macro}Leggett, A. J. Testing the limits of quantum mechanics:
motivation, state of play, prospects. J. Phys.: Condens. Matter \textbf{14,
}R415-R451 (2002).

\bibitem{Leggett2008}Leggett, A. J. Realism and the physical world.
\textit{Rep. Prog. Phys.} \textbf{71}, 022001-6 (2008).

\bibitem{transmon_th} Koch, J. \textit{et al.} Charge-insensitive
qubit design derived from the Cooper pair box. \textit{Phys. Rev.}
A \textbf{76}, 042319 (2007).

\bibitem{transmon_exp} Schreier, J.A. Suppressing charge noise decoherence
in superconducting charge qubits. \textit{et al.}, \textit{Phys. Rev.}
B \textbf{77}, 180502 (2008).

\bibitem{Gambetta}Gambetta, J. \textit{et al.} Quantum trajectory
approach to circuit QED: Quantum jumps and the Zeno effect. \textit{Phys.
Rev.} A \textbf{77}, 012112 (2008).

\bibitem{Schuster}Schuster, D.I. \textit{et al.} ac Stark Shift and
Dephasing of a Superconducting Qubit Strongly Coupled to a Cavity
Field. \textit{Phys. Rev. Lett.} \textbf{94}, 123602 (2004).

\bibitem{MisraSudarshan}Misra, B. \& Sudarshan, E.C.G. The Zeno's
paradox in quantum theory. \textit{J.} \textit{Math. Phys. Sci.} \textbf{18},
756-763 (1977).

\bibitem{Itano}Itano, W. M., Heinzen, D. J., Bollinger, J. J. and
Wineland, D. J. Quantum Zeno effect. \textit{Phys. Rev.} A \textbf{41},
2295-2300 (1990).

\bibitem{Bernu}Bernu, J. \textit{et al.} Freezing Coherent Field
Growth in a Cavity by the Quantum Zeno Effect. \textit{Phys. Rev.
Lett.} \textbf{101}, 180402 (2008).

\bibitem{Goan}Goan, H.S. \& Milburn, G.J. Dynamics of a mesoscopic
charge quantum bit under continuous quantum measurement. \textit{Phys.
Rev.} B \textbf{64}, 235307 (2001).

\bibitem{Shnirman}Shnirman, A., Mozyrsky, D., \& Martin, I. Output
spectrum of a measuring device at arbitrary voltage and temperature.
\textit{Europhys. Lett.} \textbf{67}, 840-846 (2004).

\bibitem{Torrey}Torrey, H.C. Transient Nutations in Nuclear Magnetic
Resonance. \textit{Phys. Rev.} \textbf{76}, 1059-1068 (1947).

\bibitem{Ilichev}Il'ichev, E. \textit{et al.} Continuous Monitoring
of Rabi Oscillations in a Josephson Flux Qubit. \textit{Phys. Rev.
Lett.} \textbf{91}, 097906 (2003).

\bibitem{Deblock}Deblock, R., Onac, E., Gurevich, L., \& Kouwenhoven,
L. P. Detection of Quantum Noise from an Electrically Driven Two-Level
System. \textit{Science} \textbf{301}, 203-206 (2003).

\bibitem{Manassen}Manassen, Y., Hamers, R. J., Demuth, J. E., \&
Castellano Jr, A. J. Direct observation of the precession of individual
paramagnetic spins on oxidized silicon surfaces. \textit{Phys. Rev.
Lett.} \textbf{62}, 2531-2534 (1989).

\bibitem{Qlim_amplifier1}Castellanos-Beltran, M. A., Irwin, K. D.,
Hilton, G.C., Vale L. R., \& Lehnert, K. W. Amplification and squeezing
of quantum noise with a tunable Josephson metamaterial. \textit{Nature
Physics} \textbf{4}, 929-931 (2008).

\bibitem{Qlim_amplifier2}Bergeal, N. \textit{et al.} Analog information
processing at the quantum limit with a Josephson ring modulator. arXiv:cond-mat/0805.3452. 

\end{thebibliography}

\begin{thebibliography}{S3}
\bibitem[S1]{qotoolbox} Tan, S.M. A Computational Toolbox for Quantum
and Atomic Optics, available online http://www.qo.phy.auckland.ac.nz/qotoolbox.html. 

\bibitem[S2]{MilburnWalls}Walls, D.F. \& Milburn, G. \textit{Quantum
Optics}, (Springer ed., Heidelberg 2008), chap. 6.

\bibitem[S3]{Ithier}Ithier, G. \textit{et al.} Decoherence in a superconducting
quantum bit circuit. \textit{Phys. Rev.} B \textbf{72}, 134519 (2005).
\end{thebibliography}
\end{document}